\documentclass[pdfusetitle,aps,reprint,prl,twocolumn,superscriptaddress,preprintnumbers]{revtex4-1}
\pdfoutput=1
\usepackage{amssymb}
\usepackage{amsmath}
\usepackage{epsfig}
\usepackage{hyperref}
\usepackage{breakurl}
\usepackage{multirow}
\usepackage{array}
\usepackage{bm}
\usepackage{color}
\usepackage{tikz}
\usepackage[utf8]{inputenc}
\usetikzlibrary{decorations.markings}

\makeatletter
\def\simgt{\mathrel{\lower2.5pt\vbox{\lineskip=0pt\baselineskip=0pt
           \hbox{$>$}\hbox{$\sim$}}}}
\def\simlt{\mathrel{\lower2.5pt\vbox{\lineskip=0pt\baselineskip=0pt
           \hbox{$<$}\hbox{$\sim$}}}}

\def\sectionskip{\vskip .2 cm}
\def\tfrac#1#2{{\textstyle \frac{#1}{#2}}}
\makeatother

\def\eqn#1{Eq.~\eqref{#1}}

\def\spa#1.#2{\left\langle#1\,#2\right\rangle}
\def\spb#1.#2{\left[#1\,#2\right]}
\def\sand#1.#2.#3{%
\left\langle#1{\vphantom1}\right|{#2}\left|#3\right]}%
\def\sandmp#1.#2.#3{%
\left\langle#1{\vphantom1}\right|{#2}\left|#3\right]}%
\def\sandpm#1.#2.#3{%
\left[#1{\vphantom1}\right|{#2}\left|#3\right\rangle}%
\def\sandmm#1.#2.#3{%
\left\langle#1{\vphantom1}\right|{#2}\left|#3\right\rangle}%
\def\sandpp#1.#2.#3{%
\left[#1{\vphantom1}\right|{#2}\left|#3\right]}%

\renewcommand{\imath}{\mathrm{i}}

\def\nn{\nonumber}
\def\Section#1{\noindent {\it #1}}
\newcommand{\be}{\begin{equation}}
\newcommand{\ee}{\end{equation}}

\newcommand{\pol}{\varepsilon}
\newcommand{\scaling}[1]{\scalebox{0.8}{$#1$}}
\newcommand{\scalingb}[1]{\scalebox{0.78}{$#1$}}
\newcommand{\noscaling}[1]{\scalebox{1.0}{$#1$}}

\def\H{{H_2}}
\def\S{{\mathbb S}}

\begin{document}

\title{Binary Dynamics Through the Fifth Power of Spin at ${\cal O}(G^2)$}

\author{Zvi Bern}
\affiliation{
Mani L. Bhaumik Institute for Theoretical Physics,
University of California at Los Angeles,
Los Angeles, CA 90095, USA}
\author{Dimitrios Kosmopoulos}
\affiliation{
Mani L. Bhaumik Institute for Theoretical Physics,
University of California at Los Angeles,
Los Angeles, CA 90095, USA}
\author{Andr\'es Luna}
\affiliation{
Niels Bohr International Academy,
Niels Bohr Institute, University of Copenhagen,
Blegdamsvej 17, DK-2100, Copenhagen \O , Denmark}
\author{Radu Roiban}
\affiliation{Institute for Gravitation and the Cosmos,
Pennsylvania State University,
University Park, PA 16802, USA}
\author{Fei Teng}
\affiliation{Institute for Gravitation and the Cosmos,
Pennsylvania State University,
University Park, PA 16802, USA}

\begin{abstract}

We use a previously developed scattering-amplitudes-based framework
for determining two-body Hamiltonians for generic binary systems with
arbitrary spin $S$.   By construction this formalism bypasses
difficulties with unphysical singularities or higher-time derivatives.
This framework has been previously used to obtain the exact velocity
dependence of the $\mathcal O(G^2)$ quadratic-in-spin two-body Hamiltonian.  
We first evaluate the $S^3$ scattering angle and two-body Hamiltonian at this
order in $G$, including not only all operators corresponding to the usual worldline
operators, but also an additional set due to an interesting subtlety.  We then evaluate
$S^4$ and $S^5$ contributions at $\mathcal O(G^2)$ which we confirm
by comparing against aligned-spin results.  
We conjecture that a certain shift symmetry together with a constraint on the high-energy
growth of the scattering amplitude specify the Wilson coefficients for the Kerr black hole to all orders
in the spin and confirm that they reproduce the previously-obtained results through $S^4$.
\end{abstract}

\maketitle

\sectionskip
\Section{Introduction.}
The landmark detection of gravitational waves by the LIGO/Virgo
collaboration~\cite{LIGOScientific:2016aoc,*LIGOScientific:2017vwq} heralds an era of remarkable discoveries in
astronomy, cosmology and perhaps even particle physics.  
The fundamental issues addressed in this paper, leading to the
identification of a new shift symmetry and of a subtlety in the
counting of independent interactions, lay a foundation for future
phenomenological applications.
While current detectors are not sufficiently sensitive, the remarkable precision of forthcoming 
detectors~\cite{Punturo:2010zz,*LISA:2017pwj,*Reitze:2019iox} 
will demand equally precise theoretical predictions which include detailed properties of gravitational-wave sources including their spin~\cite{Blanchet:2013haa}.

The post-Minkowskian (PM)
framework~\cite{Bertotti:1956pxu, *Kerr:1959zlt, *Bertotti:1960wuq,
  *Portilla:1979xx, *Westpfahl:1979gu, *Portilla:1980uz, *Bel:1981be,
  *Westpfahl:1985tsl, *Ledvinka:2008tk, *Damour:2016gwp}, which 
  resums the velocity expansion present in the post-Newtonian
approach~\cite{Barker:1970zr, *Barker:1975ae, *Kidder:1992fr, *Kidder:1995zr, 
*Blanchet:1998vx, *Tagoshi:2000zg, 
Porto:2005ac, *Faye:2006gx, *Blanchet:2006gy, *Damour:2007nc, *Steinhoff:2007mb, *Levi:2008nh,
*Steinhoff:2008zr, *Steinhoff:2008ji, *Marsat:2012fn,*Hergt:2010pa, 
*Porto:2010tr, *Levi:2010zu, *Porto:2010zg,*Levi:2011eq, 
*Porto:2012as, *Hergt:2012zx, *Bohe:2012mr,
*Hartung:2013dza, *Marsat:2013wwa, *Levi:2014gsa,
*Vaidya:2014kza, *Bohe:2015ana, *Bini:2017pee, *Siemonsen:2017yux,
Porto:2006bt, *Porto:2007tt, *Porto:2008tb,*Porto:2008jj,
Levi:2014sba,
Levi:2015msa,
Levi:2015uxa, *Levi:2015ixa, *Levi:2016ofk,  *Levi:2019kgk, *Levi:2020lfn,
Levi:2020kvb, *Levi:2020uwu, *Kim:2021rfj,
Maia:2017gxn, *Maia:2017yok, *Cho:2021mqw, *Cho:2022syn} has
been previously applied to the spinning two-body problem~\cite{
   Bini:2017xzy, *Bini:2018ywr,
   Maybee:2019jus, *Guevara:2019fsj,
   Chung:2020rrz,
   Guevara:2017csg, *Vines:2018gqi,*Damgaard:2019lfh, *Aoude:2020onz,
   Vines:2017hyw,  
   Guevara:2018wpp, *Chung:2018kqs,
   Chung:2019duq,
   Bern:2020buy,
   Kosmopoulos:2021zoq,
   Liu:2021zxr,
   Jakobsen:2021lvp,
   Jakobsen:2021zvh,
   Chen:2021qkk}.
In this paper we approach it with amplitudes methods.

Vines~\cite{Vines:2017hyw} obtained the energy-momentum tensor of a Kerr black hole at~${\cal O}(G)$ with the
full spin and velocity dependence and derived the corresponding two-body Hamiltonian.  
This stress tensor was shown~\cite{Guevara:2018wpp, *Chung:2018kqs, *Guevara:2019fsj, *Chung:2019duq}
to be equivalent with the minimal amplitudes of Ref.~\cite{Arkani-Hamed:2017jhn},
which were used~\cite{Chung:2020rrz} to recover
the Hamiltonian of Ref.~\cite{Vines:2017hyw}.  
At $\mathcal{O}(G^2)$, a PM spin-orbit Hamiltonian
is known~\cite{Bini:2017xzy,*Bini:2018ywr}. There has also been
progress at this order on obtaining PM higher-spin interactions~\cite{
  Guevara:2018wpp,*Chung:2018kqs,*Guevara:2019fsj,*Chung:2019duq,
  Guevara:2017csg,*Vines:2018gqi,*Damgaard:2019lfh,*Aoude:2020onz},
including the complete quadratic-in-spin interactions for the inspiral
phase of generic compact objects~\cite{Bern:2020buy,
  Kosmopoulos:2021zoq, Liu:2021zxr}. Recently, quartic-in-spin results
have been given for binary Kerr black holes~\cite{Chen:2021qkk} at $\mathcal{O}(G^2)$. 
The scattering angle for generic spinning bodies at the quadratic-in-spin level and at $\mathcal{O}(G^3)$, including radiation effects, 
was recently given in~\cite{Jakobsen:2022fcj}.

High orders in the spin bring a number of subtleties.  Among them is 
the complete categorization of all independent interactions.  
In the  worldline-effective-field-theory formalism~\cite{
Porto:2005ac, *Faye:2006gx, *Blanchet:2006gy, *Damour:2007nc, *Steinhoff:2007mb, *Levi:2008nh,
*Steinhoff:2008zr, *Steinhoff:2008ji, *Marsat:2012fn,*Hergt:2010pa, 
*Porto:2010tr, *Levi:2010zu, *Porto:2010zg,*Levi:2011eq, 
*Porto:2012as, *Hergt:2012zx, *Bohe:2012mr,
*Hartung:2013dza, *Marsat:2013wwa, *Levi:2014gsa,
*Vaidya:2014kza, *Bohe:2015ana, *Bini:2017pee, *Siemonsen:2017yux,
Porto:2006bt, *Porto:2007tt, *Porto:2008tb,*Porto:2008jj,
Levi:2014sba,
Levi:2015msa,
Levi:2015uxa, *Levi:2015ixa, *Levi:2016ofk,  *Levi:2019kgk, *Levi:2020lfn,
Levi:2020kvb, *Levi:2020uwu, *Kim:2021rfj,
Maia:2017gxn, *Maia:2017yok, *Cho:2021mqw, *Cho:2022syn} this is achieved by 
eliminating all Lagrangian terms with higher time derivatives~\cite{Schafer:1984mr, *Damour:1990jh, Levi:2014sba}. 
The amplitudes-based approaches using massive spinor
helicity~\cite{Arkani-Hamed:2017jhn} can introduce unphysical
singularities beyond the quartic-in-spin order~\cite{Chung:2019duq, Chen:2021qkk}.  
At spin-$5/2$ and beyond, Compton amplitudes free of such singularities were constructed in Refs.~\cite{Chiodaroli:2021eug,*Falkowski:2020aso,Aoude:2022trd}.
With a local Lagrangian starting point, the amplitudes-based formalism
of Ref.~\cite{Bern:2020buy} bypasses these issues to all orders in
spin. This formalism has been tested for quadratic in spin
contributions at $\mathcal O(G^2)$ in Refs.~\cite{Bern:2020buy,
Kosmopoulos:2021zoq} and confirmed using the
worldline~\cite{Liu:2021zxr} and
worldline-quantum-field-theory~\cite{Jakobsen:2021zvh} formalisms.

Amplitudes-based methods~\cite{Iwasaki:1971vb,*Iwasaki:1971iy,*Gupta:1979br,
  Donoghue:1994dn,*Bjerrum-Bohr:2002gqz,*Holstein:2008sx,*Neill:2013wsa,*Bjerrum-Bohr:2013bxa,*Cristofoli:2019neg,*Bjerrum-Bohr:2019kec,*Brandhuber:2021eyq,
  Bjerrum-Bohr:2018xdl,
  Cheung:2018wkq, Kosower:2018adc} established the state of the art in the PM
expansion
by producing the first conservative spinless two-body Hamiltonian at $\mathcal O(G^3)$ 
and $\mathcal O(G^4)$~\cite{Bern:2019nnu,*Bern:2019crd,*Bern:2021dqo,*Bern:2021yeh},
with various aspects confirmed in a number of
studies~\cite{Blumlein:2020znm,*Cheung:2020gyp,*Bini:2020wpo,*Bini:2020nsb,*Bini:2021gat,*Kalin:2020fhe,*Blumlein:2021txj,*Dlapa:2021npj,*Dlapa:2021vgp}.
Such methods led to new results that include spin~\cite{Cachazo:2017jef,Chung:2018kqs,Guevara:2018wpp,Guevara:2019fsj,Chung:2019duq,
  Vines:2018gqi,Damgaard:2019lfh,Aoude:2020onz,Chen:2021qkk} and tidal
effects~\cite{Bini:2020flp,*Kalin:2020mvi,*Cheung:2020sdj,*Haddad:2020que,*Kalin:2020lmz,*Bern:2020uwk,*Cheung:2020gbf,*Aoude:2020ygw}.
They also led to the discovery of new structures such as the double-copy relation between gauge and gravity
theories~\cite{Kawai:1985xq,*Bern:1998sv,*Bern:2008qj,*Bern:2010ue,*Bern:2019prr}
and a conjecture that a single scalar function---the eikonal phase---determines both spinning and spinless classical observables~\cite{Bern:2020buy}.

Here we use the amplitudes-based field-theory formalism of Ref.~\cite{Bern:2020buy} and a Lagrangian containing classes of operators through quintic order in spin
to derive two-body interaction Hamiltonian up to ${\cal O}(S^5)$  for black holes and more general compact objects.  To obtain a
complete description through a given power of spin one would need to
systematically add all operators that give independent contributions
to the physical observables up to power of spin of interest.

The Hamiltonian at ${\cal O}(G)$ for binary Kerr black holes is known to all orders in spin~\cite{Vines:2017hyw}.
Ref.~\cite{Chen:2021qkk} obtained a binary Hamiltonian at $\mathcal O(G^2)$ through $S^4$ from amplitudes expected to describe Kerr black holes.
Explicit results through fifth order in spin, a symmetry of the ${\cal O}(G)$ Hamiltonian 
of Ref.~\cite{Vines:2017hyw} and an additional assumption lead us to formulate a conjecture 
for the structure of the two-body Hamiltonian for Kerr black holes to all orders in spin
and determine the Hamiltonian at $S^5$.   
We consider a variety of independent interactions sufficient to illustrate the features of this conjecture, 
and leave for the future a systematic study of all possible interactions.

\sectionskip
\Section{Review of Formalism.}
In the classical limit of a scattering process, the momentum transfer $\pmb{q}$ is much smaller than 
the rest mass, $|\pmb{q}|\ll m$, while the rest-frame spin is large, $\pmb{q}\cdot\pmb{S}/m\sim\mathcal{O}(1)$.
Products of classical spin tensors are related~\cite{Bern:2020buy} to symmetric products of Lorentz generators, 
\begin{align}\label{SpinEval}
	&  \pol(p_1)\{ M^{a_1 b_1}M^{a_2 b_2} \ldots M^{a_j b_j}\} \pol(p_2) \\
	& \;
	=  S(p_1)^{a_1 b_1} S(p_1)^{a_2b_2}\cdots S(p_1)^{a_j b_j} \,
	\pol(p_1) \cdot \pol(p_2) + \cdots \,,\nonumber
\end{align}
where $\{ \ldots \}$ indicates the symmetric product, $\pol(p_1)$ and $\pol(p_2)$ are boosted spin coherent states describing the incoming and outgoing
polarization tensors of a higher-spin particle, and the ellipsis stand for subleading terms in the classical limit. 
Antisymmetric combinations of $M^{ab}$ are simplified using the Lorentz algebra. The spin tensor is obtained by boosting the one in the rest frame, 
so it respects the covariant spin supplementary condition (SSC) $p_a S^{ab}=0$. 
The classical spin vector follows from $S^{ab} \equiv -\epsilon^{a b cd} p_c S_{d}/m$, which follows by boosting the analogous rest-frame relation.

We work in a field-theory framework with a Lagrangian that simply gives a covariantization of all the spin-induced
multipole moments. Causality-based no-go theorems for the quantum consistency of higher-spin
interactions~\cite{Camanho:2014apa, Afkhami-Jeddi:2018apj} are avoided by interpreting it
as an effective theory, valid only at sufficiently large impact parameter, i.e. only in the classical regime. 
See Ref.~\cite{Chen:2021bvg} for a connection between resolvability of the time delay and the range 
of validity of an EFT. 
We first describe operators that have on-shell three-point vertices.
Following Ref.~\cite{Bern:2020buy} we separate the Lagrangian into a minimal and non-minimal part, ${\cal L}={\cal L}_\text{min} + {\cal L}_{\text{non-min}}$.
The former is
\begin{align}
{\cal L}_\text{min} &=-R + \frac{1}{2} \eta^{ab} \nabla_a\phi_s\nabla_b\phi_{s}
- \frac{1}{2}m^2\phi_s\phi_{s} \ ,
%\nn \\
\label{Ls}
\end{align}
where we use only tangent-space indices.
We take the higher-spin field $\phi_s$ to be in a real representation of the Lorentz group.  We do not require it to be transverse, so this representation is
reducible and contains spins ranging from 0 to $s$~\cite{Bern:2020buy}.
The covariant derivative is $\nabla_{c}\phi_s\equiv e_c^\mu(\partial_{\mu}\phi_s+\frac{i}{2}\omega_{\mu ab}M^{ab}\phi_s)$, 
where $e_c^\mu$ is the (inverse) vielbein, $\omega_{\mu ab}$ is the spin connection, and $M^{ab}$ 
are Lorentz generators in this 
representation. 

In the non-minimal Lagrangian, we consider two classes of linear-in-curvature operators and also selected curvature-square operators, $\mathcal{L}_\text{non-min}=\mathcal{L}_C+\mathcal{L}_H+\mathcal{L}_{R^2}$. More generally, there are infinite sequences of additional operators to consider, including those with different index contractions or higher powers of the Riemann tensor.  In general, the coefficients of these operators need to be matched to either theoretically or experimentally determined values, with coefficients being particularly simple for black holes.
The first family of linear-in-curvature operators is
\begin{align}
\label{Lnonmin}
{\cal L}_C & = \sum_{n=1}^{\infty} \frac{\left(-1\right)^n}{\left(2n\right)!}
\frac{C_{\textrm{ES}^{2n}}}{m^{2n}} \nabla_{f_{2n}}\ldots \nabla_{f_3} R_{a f_1 b f_2 } \nn \\
& \hskip 1.1 cm \null \times 
\nabla^a{\phi}_s\,  \S^{(f_1}\S^{f_2} \dots \S^{f_{2n})} \nabla^b \phi_s
\\
&\null -\sum_{n=1}^{\infty} \frac{\left(-1\right)^n}{\left(2n{+}1\right)!}
\frac{C_{\textrm{BS}^{2n+1}}}{m^{2n+1}}  \nabla_{f_{2n+1}}\ldots \nabla_{f_3} \widetilde{R}_{(a|f_1|b)f_2} \nn \\
& \hskip 1.1 cm \null \times 
\nabla^a{\phi}_s  \S^{(f_1} \S^{f_2} \dots \S^{f_{2n+1})}  \nabla^b\phi_s
\,,
\nonumber
\end{align}
where $\S^{a}\equiv \frac{-i}{2m}\epsilon^{a b cd}M_{cd}\nabla_{b}$
is the Pauli-Lubanski vector, and $\widetilde{R}_{abcd}\equiv\frac{1}{2}\epsilon_{abij}R^{ij}{}_{cd}$ is the dual Riemann tensor.
The operators in Eq.~\eqref{Lnonmin} are in one-to-one correspondence with the non-minimal operators 
in Ref.~\cite{Levi:2015msa}. 

The second family of linear-in-curvature operators we include is given by~\footnote{By using Bianchi identities, we can write the $H_2$ operator as $\frac{\H}{8} R_{abcd} \, \phi_s M^{ab} M^{cd} \phi_{s}$. This term is sometimes considered as part of the minimal Lagrangian.},
\begin{align}\label{Hs}
\mathcal{L}_{H} & =  -\sum_{n=1}^{\infty}\frac{(-1)^n(2n{-}1)}{(2n)!(2n{+}1)}\frac{H_{2n}}{m^{2n-2}}\nabla_{f_{2n}}\ldots\nabla_{f_3}R^{(a}{}_{f_1}{}^{b)}{}_{f_2}\nonumber\\
& \hskip 1.1 cm  \times\phi_sM_{a}{}^{(f_1}M_{b}{}^{f_2}\S^{f_3}\ldots\S^{f_{2n})}\phi_s\nonumber\\
&+\sum_{n=1}^{\infty}\frac{(-1)^n n}{(2n{+}1)!(n{+}1)}\frac{H_{2n+1}}{m^{2n-1}}\nabla_{f_{2n+1}}\ldots\nabla_{f_3}\widetilde{R}^{(a}{}_{f_1}{}^{b)}{}_{f_2}\nonumber\\
& \hskip 1.1 cm \null \times\phi_sM_{a}{}^{(f_1}M_{b}{}^{f_2}\S^{f_3}\ldots\S^{f_{2n+1})}\phi_s\, .
\end{align}
The normalization is chosen such that, upon using external spin tensors satisfying the
    covariant SSC, the three-point amplitudes depend only on
$C_{2n}\!\equiv\! C_{\textrm{ES}^{2n}}{+}H_{2n}$ and $C_{2n+1}\!\equiv\! C_{\textrm{BS}^{2n+1}}{+}H_{2n+1}$.
Comparison with Ref.~\cite{Vines:2017hyw} fixes $C_n=1$ for all $n\ge 2$ for a Kerr black hole. 
Because it relies on the SSC rather than the equations of motion, the equality of the three-point on-shell matrix 
elements of the operators in $\mathcal{L}_C$ and $\mathcal{L}_{H}$ {\em does not imply} that their difference is
a local higher-curvature operator. As we will see,  $\mathcal{L}_C$ and $\mathcal{L}_{H}$ contribute 
differently to the classical gravitational Compton amplitude, so both need to be included in an EFT.

Operators with higher powers of the Riemann tensor and its derivatives first contribute at $S^4$ and can also encode tidal effects.
Our conjectured structure of the two-body Hamiltonian of Kerr black holes cannot be implemented unless such terms are included starting at $S^5$ so as to cancel poor high-energy behavior.
For our conjectured QFT definition of the Kerr black hole,
it is sufficient to include:
\begin{widetext}
	\begin{align}\label{selectedRsq}
		&\mathcal{L}_{R^2}=\frac{1}{1800m^7}(E_1R_{af_1bf_2}\nabla_{f_5}\widetilde{R}_{cf_3df_4}+E_2\nabla_{f_5}R_{af_1bf_2}\widetilde{R}_{cf_3df_4})\nabla^{(a}\nabla^{c)}\phi_s \S^{(f_1}\S^{f_2}\S^{f_3}\S^{f_4}\S^{f_5)}\nabla^{(b}\nabla^{d)}\phi_s\nonumber\\
		&\quad+\frac{1}{1800m^5}(E_3R_{af_1bf_2}\nabla_{f_5}\widetilde{R}_{cf_3df_4}+E_4\nabla_{f_5}R_{af_1bf_2}\widetilde{R}_{cf_3df_4}) \nabla^c\phi_sM^{a(f_1}M^{|b|f_2}\S^{f_3}\S^{f_4}\S^{f_5)}\nabla^d\phi_s\\
		&\quad+\frac{1}{1800m^7}(2E_5R_{eabf_1}\nabla_{f_2}\widetilde{R}^{e}{}_{cdf_3}+E_6R_{aebf_1}\nabla^{e}\widetilde{R}_{cf_2df_3}+E_7\nabla_{f_2}R_{eabf_1}\widetilde{R}^{e}{}_{cdf_3})\nabla^{(a}\nabla^{c)}\phi_s\S^{m}\S_{m}\S^{(f_1}\S^{f_2}\S^{f_3)}\nabla^{(b}\nabla^{d)}\phi_s\,,\nonumber
	\end{align}
\end{widetext}
where $E_1,\ldots,E_7$ are Wilson coefficients.  For generic compact objects one would need to include 
$R^2$ operators with independent matrix elements under the SSC.

Having specified the Lagrangian $\mathcal{L} = \mathcal{L}_\text{min}+\mathcal{L}_\text{non-min}$, the four-point Compton amplitude is straightforwardly obtained using Feynman rules.
The generalized unitarity method~\cite{Bern:1994zx,*Bern:1994cg} then gives the one-loop integrand with four external higher-spin states.
The relevant generalized cut at $\mathcal O(G^2)$ is 
\begin{align}
	\textrm{Cut}_{\textrm{t-channel}}=\raisebox{-0.6cm}{\includegraphics[scale=.4]{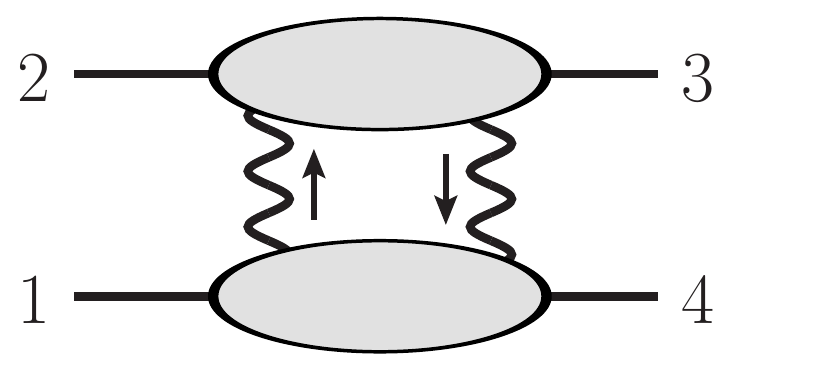}}\,,
\end{align}
where the blobs represent on-shell gravitational Compton amplitudes.
By adjusting terms that vanish on shell, the Compton amplitudes can be chosen to satisfy generalized gauge invariance, 
so the physical-state projectors are manifestly Lorentz invariant and independent of reference momenta~\cite{Kosmopoulos:2020pcd}.
Apart from inclusion of higher powers of the spin vector, the construction of the integrand follows the 
discussion in Ref.~\cite{Bern:2020buy} so we will not detail it here.
We then use FIRE~\cite{Smirnov:2008iw,*Smirnov:2019qkx} as well as Forde's method~\cite{Forde:2007mi,*Kilgore:2007qr,*Badger:2008cm} to extract the 
coefficients of the scalar triangle integrals which determine the classical amplitude. 

\sectionskip
\Section{Results.}
In general the amplitude is a sum of scalar box, triangle and bubble-integral 
contributions. The box integrals correspond to iteration of lower-order terms 
and carry no new information.  The bubble integrals contain purely quantum information 
and are hence dropped. The new classical information at this order is encoded in the 
triangle integrals and their coefficients~\cite{Bjerrum-Bohr:2018xdl,Cheung:2018wkq}, whose structure is
\begin{equation}
{\mathcal M}^{\bigtriangleup + \bigtriangledown} = 
  \frac{2 \pi^2 G^2 \pol_1 \cdot \pol_4 \pol_2 \cdot \pol_3} {\sqrt{-q^2}}
   \sum_n \sum_i \alpha^{(n,i)} {\mathcal O}^{(n,i)}\, .
\label{Mtri}
\end{equation}
Here $n$ is the number of spin vectors and the index $i$ labels their independent contractions. 
For brevity, we include here explicitly only the $S_1^n S_2^0$  terms with $n=2$ through $n=5$. 
The relevant operators ${\cal O}^{(n,i)}$ are given in Table~\ref{table:SpinStructures}, while an ancillary file
contains the full amplitudes~\cite{anc}. We use the shorthand notation ${\cal E}_j= -i\epsilon^{\mu\nu\rho\tau} u_{1\mu}u_{2\nu}q_\rho a_{j\tau}$, 
as well as $u_i^\mu=p_i^\mu/m_i$, $a_i^\mu=S_i^\mu/m_i $ and $\sigma=u_1\cdot u_2$.

%%%%%%%%%%%%%%%  TABLE %%%%%%%%%%%%%%%%%%%%%%%
\begin{table}[tb]
	\setlength{\tabcolsep}{3.5pt} 
	\setlength\extrarowheight{3.5pt} 
	\begin{tabular}{c||l|c||l|c||l|c}
		\multirow{1}{*}{$\null$} & $i$ & $\null$ & $i$ & $\null$ & $i$ & $\null$ \\ \hline
		\multirow{1}{*}{$\mathcal{O}^{(2,i)}$} %& 1 & 2 & 3 \\   \cline{2-4} 
		& $1$ & $\mathcal{E}_1^2$ & $2$ & $q^2(u_2\!\cdot\! a_1)^2$ & $3$ & $(q\!\cdot\! a_1)^2$ \\[2pt] \hline
		\multirow{2}{*}{$\mathcal{O}^{(4,i)}$} % & 1 & 2 & 3 \\  \cline{2-4} 
		& $1$ & $\mathcal{E}_1^4$ & $2$ & $q^2(u_2\!\cdot\! a_1)^2\mathcal{E}_1^2$ & $3$ & $q^4(u_2\!\cdot\! a_1)^4$ \\[2pt] \cline{2-7} 
		%& 4 & 5 & 6 \\ \cline{2-4} 
		& $4$ & ($q\!\cdot\!a_1)^2\mathcal{E}_1^2$ & $5$ & $q^2(q\!\cdot\! a_1)^2(u_2\!\cdot\! a_1)^2$ & $6$ & $(q\!\cdot\! a_1)^4$ \\ 
	\end{tabular}	
	\caption{The independent $S_1^2 S_2^0$ and $S_1^4S_2^0$
          structures are given in the table. The $S_1^3S_2^0$ and
          $S_1^5 S_2^0$ structures follow from these via
          $\mathcal{O}^{(3,i)}=\mathcal{E}_1\mathcal{O}^{(2,i)}$ and
          $\mathcal{O}^{(5,i)}=\mathcal{E}_1\mathcal{O}^{(4,i)}$. 
            }
  \label{table:SpinStructures}
\end{table}
%%%%%%%%%%%%%%%%%%%%%%%%%%%%%%%%%%%%%%%%%%%%%

We parametrize the coefficients $\alpha^{(n,i)}$ 
%of the spin structures 
in \eqn{Mtri} as
\begin{align}
\alpha^{(2,i)}&=
%%%%% begin : alphaPaper[2,i]
\frac{m_1^2m_2^2}{16(-1+\sigma^2)^2}(\gamma^{(2,i)}m_1+\delta^{(2,i)}m_2)
%%%%% end : alphaPaper[2,i]
\,,\nonumber\\  
\alpha^{(3,i)}&=
%%%%% begin : alphaPaper[3,i]
\frac{m_1^2 m_2^2 \sigma}{8(-1+\sigma^2)^2} (\gamma^{(3,i)}m_1 + 2\delta^{(3,i)}m_2 )
%%%%% end : alphaPaper[3,i]
\,,\nonumber\\
\alpha^{(4,i)}&=
%%%%% begin : alphaPaper[4,i]
\frac{m_1^2m_2^2}{1536(-1+\sigma^2)^3}\Big(\gamma^{(4,i)}m_1+\frac{8}{5}\delta^{(4,i)}m_2\Big)
%%%%% end : alphaPaper[4,i]
\,,\nonumber\\
\alpha^{(5,i)}&=
%%%%% begin : alphaPaper[5,i]
\frac{m_1^2m_2^2\sigma}{768(-1+\sigma^2)^3}\Big(\gamma^{(5,i)}m_1+\frac{1}{75}\delta^{(5,i)}m_2\Big) 
%%%%% end : alphaPaper[5,i]
 \,,
\end{align}
where $\gamma^{(k,i)}$ and $\delta^{(k,i)}$ are polynomials in $\sigma^2$. They correspond to the probe limits $m_2\ll m_1$ and $m_1\ll m_2$, respectively. 
We list the coefficients $\gamma^{(k,i)}$ in Table~\ref{table:gamma4i} in terms of combinations $Z_{k,j}$ of  the $C_n$ Wilson coefficients, which we collect in Table~\ref{table:zkj}.

%%%%%%%%%%%%%%%%%%%%%%% TABLE %%%%%%%%%%%%%%%%%
\begin{table}[tb]
	\begin{tabular}{l|c||l|c}
		$i$ & $\gamma^{(2,i)}$ & $i$ & $\gamma^{(2,i)}$ \\[2pt] \hline
		\scaling{1} & \scaling{\vphantom{\raisebox{5pt}{A}}
%%%%% begin : gammaTable[2,1]
			7+23C_2-Z_{2,1}\sigma^2(102-95\sigma^2)
%%%%% end : gammaTable[2,1]
		} & \multirow{2}{*}{\scaling{3}} & \multirow{2}{*}{\scaling{
%%%%% begin : gammaTable[2,3]
			12Z_{2,2}(\sigma^2-1)^2(5\sigma^2-1)
%%%%% end : gammaTable[2,3]
		}} \\
		\scaling{2} & \scaling{
%%%%% begin : gammaTable[2,2]
			5-11C_2+5Z_{2,1}\sigma^2(6-7\sigma^2)
%%%%% end : gammaTable[2,2]
		} & & \\[2pt] \hline
		$i$ & \noscaling{\vphantom{\raisebox{4pt}{A}} \gamma^{(3,i)}} & $i$ & \noscaling{\gamma^{(3,i)}} \\[2pt] \hline
		\scaling{1} & \scaling{\vphantom{\raisebox{5pt}{A}}
%%%%% begin : gammaTable[3,1]
			Z_{3,1}(5-9\sigma^2)
%%%%% end : gammaTable[3,1]
		} & \multirow{2}{*}{\scaling{3}} & \multirow{2}{*}{\scaling{
%%%%% begin : gammaTable[3,3]
			4Z_{3,2}(\sigma^2-1)(5\sigma^2-3)
%%%%% end : gammaTable[3,3]
		}} \\
		\scaling{2} & \scaling{
%%%%% begin : gammaTable[3,2]
			Z_{3,1}(7\sigma^2-3)
%%%%% end : gammaTable[3,2]
		} & & \\[2pt] \hline    
		\noscaling{i} & \noscaling{\vphantom{\raisebox{5pt}{A}} \gamma^{(4,i)}} & \noscaling{i} & \noscaling{\gamma^{(4,i)}} \\[2pt] \hline
		\scaling{1} & \scaling{\vphantom{\raisebox{5pt}{A}}
%%%%% begin : gammaTable[4,1]
			44C_3+59Z_{4,2}-Z_{4,1}\sigma^2(250{-}239\sigma^2)
%%%%% end : gammaTable[4,1]
		} 
		& \scaling{4} & 
		\scaling{
%%%%% begin : gammaTable[4,4]
			12Z_{4,3}(1{-}\sigma^2)(23{-}102\sigma^2+95\sigma^4)
%%%%% end : gammaTable[4,4]
		} \\
		\scaling{2} & \scaling{
%%%%% begin : gammaTable[4,2]
			72C_3-78Z_{4,2}+Z_{4,1}\sigma^2(276{-}294\sigma^2)
%%%%% end : gammaTable[4,2]
		} & \scaling{5} & \scaling{
%%%%% begin : gammaTable[4,5]
			12Z_{4,3} (\sigma^2{-}1)(11{-}30\sigma^2{+}35\sigma^4)
%%%%% end : gammaTable[4,5]
		} \\
		\scaling{3} & \scaling{
%%%%% begin : gammaTable[4,3]
			28C_3-9Z_{4,2}+7Z_{4,1}\sigma^2(2{-}3\sigma^2)
%%%%% end : gammaTable[4,3]
		} & \scaling{6} & \scaling{
%%%%% begin : gammaTable[4,6]
			24Z_{4,4}(\sigma^2{-}1)^3(5\sigma^2{-}1)
%%%%% end : gammaTable[4,6]
		} \\[2pt] \hline
		\noscaling{i} & \noscaling{\vphantom{\raisebox{5pt}{A}} \gamma^{(5,i)}} & \noscaling{i} & \noscaling{\gamma^{(5,i)}}\\[2pt] \hline
		\scaling{1} & \scaling{\vphantom{\raisebox{5pt}{A}}
%%%%% begin : gammaPaper[5,1]
			Z_{5,1} (7-13 \sigma^2) 
%%%%% end : gammaPaper[5,1]
		}
		& \scaling{4} & \scaling{
%%%%% begin : gammaPaper[5,4]
			12 Z_{5,2}(\sigma^2 - 1)(9 \sigma^2 -5 ) 
%%%%% end : gammaPaper[5,4]
		} \\
		\scaling{2} & \scaling{
%%%%% begin : gammaPaper[5,2]
			2 Z_{5,1} (11 \sigma^2-5) 
%%%%% end : gammaPaper[5,2]
		} & \scaling{5} & \scaling{
%%%%% begin : gammaPaper[5,5]
			12 Z_{5,2} (\sigma^2 - 1)  ( 3 - 7 \sigma^2 ) 
%%%%% end : gammaPaper[5,5]
		} \\
		\scaling{3} & \scaling{
%%%%% begin : gammaPaper[5,3]
			Z_{5,1} (3 \sigma^2-1) 
%%%%% end : gammaPaper[5,3]
		} & \scaling{6} & \scaling{
%%%%% begin : gammaPaper[5,6]
			8 Z_{5,3} (\sigma^2-1)^2 ( 3 - 5\sigma^2)
%%%%% end : gammaPaper[5,6]
		} \\ 	
	\end{tabular}
	\caption{The $\gamma^{(m,i)}$ polynomials for $S_1^mS_2^0$ where the $Z_{i,j}$ are
		defined in Table~\ref{table:zkj}. }
	\label{table:gamma4i} 
\end{table}
%%%%%%%%%%%%%%%%%%%%%%%%%%%%%%%%%%%%%%%%%%

%%%%%%%%%%%%%%%%%%%%%%% TABLE %%%%%%%%%%%%%%%%%
\begin{table}[tb]
\begin{tabular}{c||c}
          $Z_{2,1}=
%%%%% begin : zPaper[2,1]
              C_2+1
%%%%% end : zPaper[2,1]
%
          $ & $Z_{2,2}=
%%%%% begin : zPaper[2,2]
              C_2-1
%%%%% end : zPaper[2,2]
%
$ \\[2pt]\hline
 	  $\vphantom{\raisebox{5pt}{A}} 
           Z_{3,1}=
%%%%% begin : zPaper[3,1]
              3C_2+C_3
%%%%% end : zPaper[3,1]
%
          $ & $Z_{3,2}=
%%%%% begin : zPaper[3,2]
              C_2-C_3
%%%%% end : zPaper[3,2]
%
          $ \\[2pt] \hline
	  $\vphantom{\raisebox{5pt}{A}}
           Z_{4,1}=
%%%%% begin : zPaper[4,1]
              3C_2^2+4C_3+C_4
%%%%% end : zPaper[4,1]
%
          $ & $Z_{4,3}=
%%%%% begin : zPaper[4,3]
              C_2^2-C_4
%%%%% end : zPaper[4,3]
%
          $ \\
 	  $  Z_{4,2}=
%%%%% begin : zPaper[4,2]
              3C_2^2+C_4
%%%%% end : zPaper[4,2]
%
          $ & $Z_{4,4}=
%%%%% begin : zPaper[4,4]
              3C_2^2-4C_3+C_4
%%%%% end : zPaper[4,4]
%
          $ \\[2pt] \hline
	  \multirow{2}{*}{$\vphantom{\raisebox{5pt}{A}}
                            Z_{5,1}=
%%%%% begin : zPaper[5,1]
            10C_2 C_3+5C_4+C_5
%%%%% end : zPaper[5,1]
%
          $} & $\vphantom{\raisebox{5pt}{A}} 
               Z_{5,2}=
%%%%% begin : zPaper[5,2]
            2C_2 C_3-C_4-C_5
%%%%% end : zPaper[5,2]
%
          $ \\
	  & $Z_{5,3}=
%%%%% begin : zPaper[5,3]
            2C_2 C_3-3C_4+C_5
%%%%% end : zPaper[5,3]
          $ 
\end{tabular}
	\caption{Useful combinations of Wilson coefficients.}
	\label{table:zkj}
\end{table}
%%%%%%%%%%%%%%%%%%%%%%%TABLE %%%%%%%%%%%%%%%%%%%%%%%%

%%%%%%%%%%%%%%%%%% TABLE %%%%%%%%%%%%%%%%%%%%%%%%%%%%%%%%%%%%%%%%%%
\begin{table}[tb]
	\begin{tabular}{c|>{\centering\arraybackslash}p{0.21\columnwidth}|>{\centering\arraybackslash}p{0.21\columnwidth}|>{\centering\arraybackslash}p{0.21\columnwidth}|>{\centering\arraybackslash}p{0.21\columnwidth}}
		$i$ & $\delta^{(2,i)}_0$ & $\delta^{(2,i)}_1$ & $\delta^{(2,i)}_2$ & $\delta^{(2,i)}_3$ \\[2pt] \hline
		\scaling{1} & \scaling{\vphantom{\raisebox{5pt}{A}}
%%%%% begin : deltaTable[0][2,1]
                        8Z_{2,1}
%%%%% end : deltaTable[0][2,1]
                } & \scaling{
%%%%% begin : deltaTable[1][2,1]
                        -(68+52C_2)
%%%%% end : deltaTable[1][2,1]
%
                } & \scaling{
%%%%% begin : deltaTable[2][2,1]
                         60Z_{2,1}
%%%%% end : deltaTable[2][2,1]
%
%
                } & \scaling{
%%%%% begin : deltaTable[3][2,1]
                         0
%%%%% end : deltaTable[3][2,1]
                } \\
		\scaling{2} & \scaling{
%%%%% begin : deltaTable[0][2,2]
                          4(3-C_2)
%%%%% end : deltaTable[0][2,2]
                } & \scaling{
%%%%% begin : deltaTable[1][2,2]
                          -12Z_{2,1}
%%%%% end : deltaTable[1][2,2]
                } & \scaling{
%%%%% begin : deltaTable[2][2,2]
                           0
%%%%% end : deltaTable[2][2,2]
                } & \scaling{
%%%%% begin : deltaTable[3][2,2]
                           0
%%%%% end : deltaTable[3][2,2]
                } \\
		\scaling{3} & \scaling{
%%%%% begin : deltaTable[0][2,3]
                           -4Z_{2,2}
%%%%% end : deltaTable[0][2,3]
                } & \scaling{
%%%%% begin : deltaTable[1][2,3]
                            68Z_{2,2}
%%%%% end : deltaTable[1][2,3]
                } & \scaling{
%%%%% begin : deltaTable[2][2,3]
                           -124Z_{2,2}
%%%%% end : deltaTable[2][2,3]
                } & \scaling{
%%%%% begin : deltaTable[3][2,3]
                            60Z_{2,2}
%%%%% end : deltaTable[3][2,3]
                }\\[2pt] \hline
		$i$ & \multicolumn{2}{c|}{ \noscaling{\vphantom{\raisebox{5pt}{A}} \delta^{(3,i)}_0}} & \multicolumn{2}{c}{ \noscaling{\delta^{(3,i)}_1}} \\[2pt] \hline
		\scaling{1} & \multicolumn{2}{c|}{\scaling{\vphantom{\raisebox{5pt}{A}}
%%%%% begin : deltaTable[0][3,1]
			3(H_2-2)H_2-(C_2-8)C_2
%%%%% end : deltaTable[0][3,1]
		}} & \multicolumn{2}{c}{\scaling{
%%%%% begin : deltaTable[1][3,1]
			C_2(2C_2-13)-2C_3-3(H_2-2)H_2
%%%%% end : deltaTable[1][3,1]
		}} \\
		\scaling{2} & \multicolumn{2}{c|}{\scaling{
%%%%% begin : deltaTable[0][3,2]
			C_2(4C_2-5)+2C_3-5(H_2-2)H_2
%%%%% end : deltaTable[0][3,2]
		}} & \multicolumn{2}{c}{\scaling{
%%%%% begin : deltaTable[1][3,2]
			5(C_2-H_2)(2-C_2-H_2)
%%%%% end : deltaTable[1][3,2]
		}} \\
		\scaling{3} & \multicolumn{2}{c|}{\scaling{
%%%%% begin : deltaTable[0][3,3]
			(5-2C_2)C_2-2C_3+(H_2-2)H_2
%%%%% end : deltaTable[0][3,3]
		}} & \multicolumn{2}{c}{\scaling{
%%%%% begin : deltaTable[1][3,3]
			2\big[C_2(3C_2-8)+4C_3-(H_2-2)H_2\big]
%%%%% end : deltaTable[1][3,3]
		}} \\[2pt] \hline
		    $ \vphantom{\raisebox{4pt}{A}}$   & \multicolumn{4}{c}{\scaling{
				\delta_2^{(3,1)}=\delta_2^{(3,2)}=
%%%%% begin : deltaTable[2][3,{1,2}]
			0
%%%%% end : deltaTable[2][3,{1,2}]
				\,, \hskip .7 cm 
				\delta_2^{(3,3)}=
%%%%% begin : deltaTable[2][3,3]
			(11-4C_2) C_2-6C_3+(H_2-2)H_2
%%%%% end : deltaTable[2][3,3]
		}}\\[2pt] \hline
	\end{tabular}
	\caption{Coefficients of the polynomials $\delta^{(2,i)}$ and $\delta^{(3,i)}$.}
	\label{table:delta3i}
\end{table}
%%%%%%%%%%%%%%%%%%%%  TABLE %%%%%%%%%%%%%%

%%%%%%%%%%%%%%%%%%%%%%%%%% TABLE %%%%%%%%%%%%%%%%%%%%
\begin{table*}[tb]
	\begin{tabular}{c|l|c|l|c}
		\multirow{3}{*}{\noscaling{\delta_0^{(4,i)}}}  & \scaling{1} & \scaling{
%%%%% begin : deltaTable[0][4,1]
			8 \big[45 C_2^2 + 5 C_3 - 5 C_2 (C_3 + 9 H_2) + 5 (H_2-1) H_3 - H_4\big]
%%%%% end : deltaTable[0][4,1]
		} & 
		\scaling{4} & 
		\scaling{
%%%%% begin : deltaTable[0][4,4]
			10 \big[ 4 C_4-5 C_3 + C_2 (36 C_2 - 11 C_3 - 24 H_2) - 5 H_3 ( H_2 -1)\big] - 4 H_4
%%%%% end : deltaTable[0][4,4]
		} \\
		&
		\scaling{2} & 
		\scaling{
%%%%% begin : deltaTable[0][4,2]
			-10 \big[36 C_2^2 - 7 C_3 - 25 C_2 C_3 - 4 C_4 - 48 C_2 H_2 + 9 (H_2-1) H_3\big] - 4 H_4
%%%%% end : deltaTable[0][4,2]
		} & 
		\scaling{5} 
		& \scaling{
%%%%% begin : deltaTable[0][4,5]
			10 \big[C_2 (9 C_2 + 10 C_3 - 15 H_2) + 4 H_2 H_3 - 2 (C_3 + C_4 + 2 H_3)\big] + 8 H_4
%%%%% end : deltaTable[0][4,5]
		} \\
		&\scaling{3} & \scaling{
%%%%% begin : deltaTable[0][4,3]
			4 H_4-10 C_4 + 15 C_2 (9 C_2 - 11 H_2)
%%%%% end : deltaTable[0][4,3]
		} & \scaling{6} & \scaling{
%%%%% begin : deltaTable[0][4,6]
			135 C_2^2 - 5 C_2 (8 C_3 + 9 H_2) - 10 \big[4 C_3 + C_4 + 4 (H_2-1) H_3\big] + 4 H_4
%%%%% end : deltaTable[0][4,6]
		} \\[2pt] \hline
		\multirow{3}{*}{\noscaling{\delta_1^{(4,i)}}}    &\scaling{1} & \scaling{\vphantom{\raisebox{6pt}{A}}
%%%%% begin : deltaTable[1][4,1]
			-10 \big[120 C_2^2 + 51 C_3 + 4 C_4 - C_2 (19 C_3 {+} 108 H_2) + 19 (H_2{-}1) H_3 - 2 H_4\big]
%%%%% end : deltaTable[1][4,1]
		} & \scaling{4} & \scaling{
%%%%% begin : deltaTable[1][4,4]
			2 \big[65 C_3-825 C_2^2 + 50 C_4 + 5 C_2 (43 C_3 {+} 99 H_2) + 35 (H_2{-}1) H_3 + 4 H_4\big]
%%%%% end : deltaTable[1][4,4]
		} \\
		&\scaling{2} & \scaling{
%%%%% begin : deltaTable[1][4,2]
			2 \big[615 C_2^2 + 85 C_3 - 50 C_4 - 5 C_2 (49 C_3 {+} 153 H_2) + 45 (H_2 {-} 1) H_3 + 8 H_4\big]
%%%%% end : deltaTable[1][4,2]
		} & \scaling{5} & \scaling{
%%%%% begin : deltaTable[1][4,5]
			10 \big[15 C_2^2 - 5 C_3 - 11 C_2 C_3 + 16 C_4 - 15 C_2 H_2 - 23 (H_2 {-} 1) H_3\big] - 28 H_4
%%%%% end : deltaTable[1][4,5]
		} \\
		&\scaling{3} & \scaling{
%%%%% begin : deltaTable[1][4,3]
			50 \big[C_3 + (H_2-1) H_3\big] -165 C_2^2 - 5 C_2 (10 C_3 - 33 H_2) - 4 H_4
%%%%% end : deltaTable[1][4,3]
		} 
		& \scaling{6} & \scaling{
%%%%% begin : deltaTable[1][4,6]
			345 C_2 H_2 -825 C_2^2 + 80 \big[6 C_3 + (H_2-1) H_3\big] - 24 H_4
%%%%% end : deltaTable[1][4,6]
		} \\[2pt] \hline
		\multirow{3}{*}{  \noscaling{\delta_2^{(4,i)}}}       & \scaling{1} & \scaling{\vphantom{\raisebox{6pt}{A}}
%%%%% begin : deltaTable[2][4,1]
			5 \big[243 C_2^2 + 78 C_3 - 14 C_2 C_3 + 14 C_4 - 201 C_2 H_2 + 30 (H_2 {-} 1) H_3\big] - 12 H_4
%%%%% end : deltaTable[2][4,1]
		} & \scaling{4} & \scaling{
%%%%% begin : deltaTable[2][4,4]
			10 \big[273 C_2^2 - 33 C_3 - 31 C_2 C_3 - 44 C_4 - 165 C_2 H_2 + (H_2 {-} 1) H_3\big] - 4 H_4
%%%%% end : deltaTable[2][4,4]
		}
		\\
		& \scaling{5} & \scaling{
%%%%% begin : deltaTable[2][4,2]
			-2 \big[525 C_2^2 - 55 C_3 + C_2 (55 C_3 - 525 H_2) - 175 (H_2 {-} 1) H_3 + 6 H_4\big]
%%%%% end : deltaTable[2][4,2]
		} & \scaling{5} & \scaling{
%%%%% begin : deltaTable[2][4,5]
			-10 \big[69 C_2^2 - 51 C_3 + 43 C_2 C_3 + 14 C_4 - 75 C_2 H_2 {-} 69 (H_2 {-} 1) H_3\big] {+} 32 H_4
%%%%% end : deltaTable[2][4,5]
		} \\
		& \scaling{3} & \scaling{
%%%%% begin : deltaTable[2][4,3]
			50 \big[(C_2-1) C_3 + H_3 - H_2 H_3\big]
%%%%% end : deltaTable[2][4,3]
		} & \scaling{6} & \scaling{
%%%%% begin : deltaTable[2][4,6]
			2 \big[\!-640 C_3 + 30 C_4 + 5 C_2 (195 C_2 + 32 C_3 - 105 H_2) + 24 H_4\big]
%%%%% end : deltaTable[2][4,6]
		} \\[2pt] \hline

		\multirow{3}{*}{\noscaling{\delta_3^{(4,i)}}}  & \scaling{1} & \scaling{\vphantom{\raisebox{6pt}{A}}
%%%%% begin : deltaTable[3][4,1]
			5 \big[16 C_3 - C_2 (57 C_2 + 16 C_3 - 57 H_2)\big]
%%%%% end : deltaTable[3][4,1]
		} & \scaling{4} & \scaling{
%%%%% begin : deltaTable[3][4,4]
			-10 \big[183 C_2^2 - 47 C_3 + 23 C_2 C_3 - 30 C_4 - 129 C_2 H_2 + 3 (H_2{-}1) H_3\big]
%%%%% end : deltaTable[3][4,4]
		} \\
		& \scaling{2} & \scaling{
%%%%% begin : deltaTable[3][4,2]
			350 \big[(C_2-1) C_3 + H_3 - H_2 H_3\big]
%%%%% end : deltaTable[3][4,2]
		} & \scaling{5} & \scaling{
%%%%% begin : deltaTable[3][4,5]
			10 \big[45 C_2^2 - 79 C_3 + 79 C_2 C_3 - 45 C_2 H_2 - 85 (H_2{-}1) H_3\big] - 12 H_4
%%%%% end : deltaTable[3][4,5]
		} \\
		&  \scaling{3}& \scaling{
%%%%% begin : deltaTable[3][4,3]
                 0
%%%%% end : deltaTable[3][4,3]
              } & \scaling{6} & \scaling{
%%%%% begin : deltaTable[3][4,6]
			-10 \big[C_2 (225 C_2 {+} 56 C_3 {-} 153 H_2) + 8 H_2 H_3 + 4 (-34 C_3 {+} 2 C_4 {-} 2 H_3 {+} H_4 ) \big]
%%%%% end : deltaTable[3][4,6]
		} \\[2pt] \hline

		\multirow{3}{*}{\noscaling{\delta_4^{(4,i)}}}  & \scaling{1} & \scaling{\vphantom{\raisebox{6pt}{A}}
%%%%% begin : deltaTable[4][4,1]
			0
%%%%% end : deltaTable[4][4,1]
		} & \scaling{4} & \scaling{
%%%%% begin : deltaTable[4][4,4]
			 10 \big[-22 C_3 + C_2 (39 C_2 + 22 C_3 - 39 H_2)\big]
%%%%% end : deltaTable[4][4,4]
		} \\
		& \scaling{2} & \scaling{
%%%%% begin : deltaTable[4][4,2]
			0
%%%%% end : deltaTable[4][4,2]
		} & \scaling{5} & \scaling{
%%%%% begin : deltaTable[4][4,5]
                        -350 \big[(C_2-1) C_3 - H_3 ( H_2 - 1 )\big]
%%%%% end : deltaTable[4][4,5]
		} \\
		&  \scaling{3}& \scaling{
%%%%% begin : deltaTable[4][4,3]
                     0
%%%%% end : deltaTable[4][4,3]
                   } & \scaling{6} & \scaling{
%%%%% begin : deltaTable[4][4,6]
		  5 \big[3 \big(-40 C_3 + 2 C_4 + C_2 (85 C_2 {+} 24 C_3 {-} 71 H_2)\big) + 8 (H_2-1) H_3\big] + 12 H_4
%%%%% end : deltaTable[4][4,6]
		} \\[2pt] \hline
		\multirow{3}{*}{\noscaling{\delta_5^{(4,i)}}}  & \scaling{1} & \scaling{\vphantom{\raisebox{6pt}{A}}
%%%%% begin : deltaTable[5][4,1]
			0
%%%%% end : deltaTable[5][4,1]
		} & \scaling{4} & \scaling{
%%%%% begin : deltaTable[5][4,4]
			0
%%%%% end : deltaTable[5][4,4]
		} \\
		& \scaling{2} & \scaling{
%%%%% begin : deltaTable[5][4,2]
			0
%%%%% end : deltaTable[5][4,2]
		} & \scaling{5} & \scaling{
%%%%% begin : deltaTable[5][4,5]
                        0
%%%%% end : deltaTable[5][4,5]
		} \\
		&  \scaling{3}& \scaling{
%%%%% begin : deltaTable[5][4,3]
                      0
%%%%% end : deltaTable[5][4,3]
                   } & \scaling{6} & \scaling{
%%%%% begin : deltaTable[5][4,6]
                  5 \big[16 C_3 - C_2 (57 C_2 + 16 C_3 - 57 H_2)\big]
%%%%% end : deltaTable[5][4,6]
		} \\[2pt] \hline
		\multirow{6}{*}{\noscaling{\delta^{(5,i)}_0}} & \scaling{1} & \multicolumn{3}{c}{\scaling{\vphantom{\raisebox{5pt}{A}} 
%%%%% begin : deltaTable[0][5,1]
				50 \big[ 39 C_4-22 C_3^2 - 72 C_3 H_2 - 9 C_4 H_2 + 24 C_2 (7 C_3 - 2 H_3) + 6 H_3^2 + 21 (H_2{-}1) H_4\big] - 69 H_5 + 2 (7 E_5 + 4 E_6 - 7 E_7 - 2 E_2 + E_4 + 
				8 E_3 + 2 E_1)
%%%%% end : deltaTable[0][5,1]
		}}
		\\[1pt]
		& \scaling{2} & \multicolumn{3}{c}{\scaling{ 
%%%%% begin : deltaTable[0][5,2]
				300 C_2 (2H_3-13 C_3) + 25 \big[128 C_3^2 - 57 C_4 + 30 C_5 + 180 C_3 H_2 + 12 C_4 H_2 + 8 H_3^2 - 66 (H_2{-}1) H_4\big] + 37 H_5 + 14 E_5 - 14 E_7 - E_2 - 31 E_4 - 13 E_3 + E_1
%%%%% end : deltaTable[0][5,2]
		} } \\[1pt]
		& \scaling{3} & \multicolumn{3}{c}{\scaling{ 
%%%%% begin : deltaTable[0][5,3]
				50 \big[14 C_3^2 -96 C_2 C_3 + 66 C_3 H_2 + 9 C_4 H_2 + 54 C_2 H_3 - 6 H_2 H_4 + 6 (C_4 + C_5 + H_4)\big] + 106 H_5 -8 E_6 + 3 E_2 - 33 E_4 - 29 E_3 - 3 E_1
%%%%% end : deltaTable[0][5,3]
		} } \\[1pt]
		& \scaling{4} & \multicolumn{3}{c}{ \scaling{
%%%%% begin : deltaTable[0][5,4]
				240 C_5 {-} 5325 C_4 {+} 1050 H_4 {+} 50 \big[42 C_3 H_2 {-} 56 C_3^2 {+} 18 C_4 H_2 {+} 6 C_2 (25 C_3 {-} 8 H_3) {+} 16 H_3^2 {-} 21 H_2 H_4\big] {+} 527 H_5 {+} 18 E_5 {+} 14 E_6 {-}34 E_7 {-} 11 E_2 {+} 19 E_4 {+} 
				47 E_3 {+} 3 E_1
%%%%% end : deltaTable[0][5,4]
		}} \\[1pt]
		& \scaling{5} & \multicolumn{3}{c}{\scaling{ 
%%%%% begin : deltaTable[0][5,5]
				660 (5 C_4 {-} 6 C_5 {-} 5 H_4) {+} 50 \big[88 C_3^2 {-} 66 C_3 H_2 {-} 111 C_4 H_2 {-} 24 H_3^2 {+} 6 C_2 (23 H_3{-}17 C_3) {+} 66 H_2 H_4\big] {+} 702 H_5 + 4 E_5 - 2 E_6 - 20 E_7 - E_2 - 61 E_4 - 
				43 E_3 - E_1
%%%%% end : deltaTable[0][5,5]
		}} \\[1pt]
		& \scaling{6} & \multicolumn{3}{c}{\scaling{
%%%%% begin : deltaTable[0][5,6]
				20 \big[30 H_4 {-} 285 C_4 {-} 48 C_5 {+} 5 (18 C_3 H_2{-}8 C_3^2 {+} 24 C_4 H_2 {+} 9 C_2 (5 C_3 {-} 2 H_3) {+} 8 H_3^2 {-} 6 H_2 H_4)\big] + 596 H_5 + 2 \big[2 E_5 + 3 E_6 - 10 E_7 - 2 E_2 - 
				7 (2 E_4 {+} E_3) + E_1\big]
%%%%% end : deltaTable[0][5,6]
		}} \\[1pt] \hline
		\multirow{6}{*}{\noscaling{\delta^{(5,i)}_1}} & \scaling{1} & \multicolumn{3}{c}{\scaling{ \vphantom{\raisebox{5pt}{A}} 
%%%%% begin : deltaTable[1][5,1]
				-75 \big[228 C_2 C_3 + 39 C_4 + 6 C_5 - 76 C_3 H_2 - 4 C_4 H_2 - 72 C_2 H_3 + 8 H_3^2 + 14 (H_2{-}1) H_4\big] + 69 H_5 -42 E_5 - 24 E_6 + 42 E_7 + 11 E_2 + 5 E_4 - 
				11 (3 E_3 {+} E_1)
%%%%% end : deltaTable[1][5,1]
		}} \\[1pt]
		& \scaling{2} & \multicolumn{3}{c}{\scaling{
%%%%% begin : deltaTable[1][5,2]
				50 \big[33 C_4 + 6 C_5 + 48 C_3 H_2 + 72 C_4 H_2 - 36 C_2 (C_3 {-} 3 H_3) - 44 H_3^2 - 36 (H_2{-}1) H_4\big] - 212 H_5 -8 (7 E_5 + 2 E_6 - 7 E_7 - E_2 - 10 E_4 - 
				3 E_3 + E_1)
%%%%% end : deltaTable[1][5,2]
		}} \\[1pt]
		& \scaling{3} & \multicolumn{3}{c}{\scaling{ 
%%%%% begin : deltaTable[1][5,3]
				-75 \big[9 C_4-84 C_2 C_3 + 2 C_5 + 44 C_3 H_2 - 4 C_4 H_2 + 40 C_2 H_3 + 8 H_3^2 + 10 (H_2{-}1) H_4\big] - 281 H_5 -14 E_5 + 8 E_6 + 14 E_7 - 3 E_2 + 75 E_4 + 
				57 E_3 + 3 E_1
%%%%% end : deltaTable[1][5,3]
		}} \\[1pt]
		& \scaling{4} & \multicolumn{3}{c}{\scaling{ 
%%%%% begin : deltaTable[1][5,4]
				60 (250 C_4 {+} 52 C_5 {-} 35 H_4) {+} 50 \big[56 C_3^2 {-}3H_2(6C_3{+}29C_4{-}14H_4) {-} 48 H_3^2 {+} 6 C_2 (35 H_3 {-} 93 C_3) \big] {-} 1054 H_5 {-} 92 E_5 {-} 66 E_6 {+} 124 E_7 {+} 39 E_2 {-} 69 E_4 {-} 
				203 E_3 {-} 17 E_1
%%%%% end : deltaTable[1][5,4]
		}} \\[1pt]
		& \scaling{5} & \multicolumn{3}{c}{\scalingb{ 
%%%%% begin : deltaTable[1][5,5]
				5 \big[3060 C_3 H_2{-}880 C_3^2 {+} 60 C_2 (9 C_3 {-} 34 H_3) {+} 3 (328 C_5 {-} 5 C_4 (169 {-} 200 H_2) {+} 80 H_3^2 {-} 730 (H_2 {-} 1) H_4)\big] {-} 1579 H_5 {-} 50 E_5  {+} 82 E_7 {+} 9 (E_2 {+} 19 E_4 {-} 2 E_6) {+} 
				67 E_3 {-} 5 E_1
%%%%% end : deltaTable[1][5,5]
		}} \\[1pt]
		& \scaling{6} & \multicolumn{3}{c}{\scaling{
%%%%% begin : deltaTable[1][5,6]
				20 \big[80 C_3^2 {+} 1095 C_4 {+} 84 C_5 {-} 120 C_3 H_2 {-} 510 C_4 H_2 {-} 160 H_3^2 {+} 60 C_2 (9 H_3 {-} 19 C_3) {+} 90 (H_2 {-} 1) H_4\big] {-} 1788 H_5 {-} 8 (5 E_5 {+} 5 E_6 {-} 11 E_7 {-} 2 E_2 {-} 17 E_4 {-} 
				5 E_3 {+} 2 E_1)
%%%%% end : deltaTable[1][5,6]
		}} \\[1pt] \hline
		\multirow{6}{*}{\noscaling{\delta^{(5,i)}_2}} & \scaling{1} & \multicolumn{3}{c}{\scaling{ \vphantom{\raisebox{5pt}{A}} 
%%%%% begin : deltaTable[2][5,1]
				50 \big[22 C_3^2 + 21 C_4 (H_2-1) - 42 C_3 H_2 + 42 C_2 (2 C_3 - H_3)\big] + 28 E_5 + 16 E_6 - 28 E_7 - 7 E_2 - 7 E_4 + 
				17 E_3 + 7 E_1
%%%%% end : deltaTable[2][5,1]
		}} \\[1pt]
		& \scaling{2} & \multicolumn{3}{c}{\scaling{
%%%%% begin : deltaTable[2][5,2]
				25 \big[153 C_4 {+} 4 (147 C_2 C_3 {-} 32 C_3^2 {-} 69 C_3 H_2 {-} 57 C_4 H_2 {-} 78 C_2 H_3 {+} 26 H_3^2) {+} 138 H_2 H_4 {-} 6 (C_5 {+} 23 H_4) {+} 7 H_5\big] {+} 42 E_5 {+} 16 E_6 {-} 42 E_7 {-} 7 E_2 {-} 49 E_4 {-} 
				11 E_3 {+} 7 E_1
%%%%% end : deltaTable[2][5,2]
		}} \\[1pt]
		& \scaling{3} & \multicolumn{3}{c}{\scaling{
%%%%% begin : deltaTable[2][5,3]
				-175 \big[4 C_3^2 + 6 C_4 (H_2-1) - 4 H_3^2 + 6 H_4 - 6 H_2 H_4 - H_5\big] +14 (E_5 - E_7 - 3 E_4 - 2 E_3) 
%%%%% end : deltaTable[2][5,3]
		}} \\[1pt] 
		& \scaling{4} & \multicolumn{3}{c}{\scalingb{ 
%%%%% begin : deltaTable[2][5,4]
				15 (70 H_4 {-} 815 C_4 {-} 224 C_5 ) {+} 50 \big[56 C_3^2 {-} 3H_2 (30C_3 {-} 40 C_4 {+} 7H_4) {+} 6 C_2 (87 C_3 {-} 38 H_3) {+} 32 H_3^2 \big] {+} 527 H_5 {+} 5( 26E_5 {+} 18 E_6 {+} 
				53 E_3 {+} 5 E_1) {-} 146 E_7 {-} 9(5 E_2 {-} 9 E_4) 
%%%%% end : deltaTable[2][5,4]
		}} \\[1pt]
		& \scaling{5} & \multicolumn{3}{c}{\scalingb{ 
%%%%% begin : deltaTable[2][5,5]
				10 \big[15C_4 (91{-} 89H_2) {-}440 C_3^2 {-} 156 C_5 {-} 2070 C_3 H_2  {+} 90 C_2 (19 C_3 {-} 3 H_3) {+} 200 (H_3^2 {+} 6 (H_2{-}1) H_4)\big] {+} 1052 H_5 {+}88 E_5 {+} 42 E_6 {-} 104 E_7 {-} 15 E_2 {-} 159 E_4 {-}
				5 E_3 {+} 13 E_1
%%%%% end : deltaTable[2][5,5]
		}} \\[1pt]
		& \scaling{6} & \multicolumn{3}{c}{\scaling{
%%%%% begin : deltaTable[2][5,6]
				4 \big[ 75 C_4 (46 H_2{-}91) {-}120 C_5 {+} 150 C_2 (65 C_3 {-} 32 H_3) {+} 50 (20 H_3^2 {+} 9 H_4 {-} 9 H_2 (2 C_3 {+} H_4)) {+} 447 H_5\big] {+} 12 (8 E_5 {+} 7 E_6 {-} 12 E_7 {-} 2 E_2 {-} 20 E_4 {-} 
				3 E_3 {+} 3 E_1)
%%%%% end : deltaTable[2][5,6]
		}} \\[1pt] \hline
		\multirow{3}{*}{\noscaling{\delta^{(5,i)}_3}} & \scaling{4} & \multicolumn{3}{c}{\scaling{\vphantom{\raisebox{5pt}{A}} 
%%%%% begin : deltaTable[3][5,4]
				-50 \big[56 C_3^2 + 51 C_4 (H_2-1) - 66 C_3 H_2 + 6 C_2 (19 C_3 - 11 H_3)\big]-56 E_5 - 38 E_6 + 56 E_7 + 17 E_2 - 31 E_4 - 
				109 E_3 - 11 E_1
%%%%% end : deltaTable[3][5,4]
		}} \\[1pt]
		& \scaling{5} & \multicolumn{3}{c}{\scaling{
%%%%% begin : deltaTable[3][5,5]
				25 (176 C_3^2 {-} 588 C_2 C_3 {-} 171 C_4 {+} 24 C_5 {+} 348 C_3 H_2 {+} 156 C_4 H_2 {+} 240 C_2 H_3 {-} 80 H_3^2 {-} 174 (H_2 {-}1) H_4 {-} 7 H_5) {-} 42 E_5 {-} 22 E_6 {+} 42 E_7 {+} 7 E_2 {+} 49 E_4 {-} 
				19 E_3 {-} 7 E_1
%%%%% end : deltaTable[3][5,5]
		}} \\[1pt]
		& \scaling{6} & \multicolumn{3}{c}{\scaling{ 
%%%%% begin : deltaTable[3][5,6]
				20 \big[585 C_4 {-} 12 C_5 {-} 30 H_4 {-} 10 (138 C_2 C_3 {+} 8 C_3^2 {-} 36 C_3 H_2 {+} 33 C_4 H_2 {-} 66 C_2 H_3 {+} 8 H_3^2 {-} 3 H_2 H_4)\big] {-} 596 H_5 {-}8 (11 E_5 {+} 9 E_6 {-} 13 E_7 {-} 2 E_2 {-} 23 E_4 {-} 
				E_3 {+} 4 E_1)
%%%%% end : deltaTable[3][5,6]
		}} \\[1pt] \hline
		\noscaling{\delta^{(5,i)}_4} & \scaling{6} & \multicolumn{3}{c}{\scaling{ \vphantom{\raisebox{5pt}{A}} 
%%%%% begin : deltaTable[4][5,6]
				100 \big[8 C_3^2 + 6 C_4 (H_2-1) - 30 C_3 H_2 + 3C_2 (23 C_3 - 10 H_3)\big] + 2 (14 E_5 + 11 E_6 - 14 E_7 - 2 E_2 - 26 E_4 + 
				E_3 + 5 E_1)
%%%%% end : deltaTable[4][5,6]
		}}
	\end{tabular}
	\caption{The coefficients of the $\delta^{(4,i)}$ and $\delta^{(5,i)}$ polynomials for the $S_1^4S_2^0$ and $S_1^5S_2^0$ amplitudes. Only nonzero coefficients of $\delta^{(5,i)}$ are given in the table.}
	\label{table:delta4i}
\end{table*}
%%%%%%%%%%%%%%%%%%%%% TABLE %%%%%%%%%%%%%%%%%%%

We parametrize the polynomials $\delta^{(k,i)}$ that govern the limit $m_1\ll m_2$
as
\begin{align}
  \delta^{(k, i)} &= \sum_{\ell=0}^{5}\delta^{(k,i)}_{\ell}\sigma^{2\ell} \,.
\end{align}
The coefficients $\delta^{(k,i)}_{\ell}$ determining $\delta^{(2,i)}$ and $\delta^{(3,i)}$ 
are given 
in Table~\ref{table:delta3i}. We note that while $\delta^{(2,i)}_{\ell}$ depends only
on the combination $C_2$ of Wilson coefficients, $\delta^{(3,i)}_{\ell}$ depends separately on 
$C_2$ and $H_2$. The coefficients of the polynomials $\delta^{(4,i)}$ and $\delta^{(5,i)}$ 
are given in Table~\ref{table:delta4i} and depend separately on $C_n$, $H_n$ and $E_n$.

For Kerr black holes, the coefficients $C_n=1$ set to zero all the $\gamma^{(k,i)}$ coefficients in the second column of Tables~\ref{table:gamma4i} and~\ref{table:zkj}.
The remaining operators in Table~\ref{table:SpinStructures} with nonzero $\gamma^{(k,i)}$ coefficients 
have, up to quantum-suppressed terms $p_i\cdot q=\pm q^2/2\rightarrow 0$, the shift symmetry 
\begin{align}
a_i^{\mu} \rightarrow a_i^{\mu} + \xi_i q^{\mu}/q^2, \quad i=1,2 \ ,
\label{symmetry}
\end{align}
where $\xi_i$ are arbitrary constants and $q^2$ was included so the classical $q$ scaling is uniform.
Inspection of the all-orders-in-spin 1PM scattering amplitude of two Kerr black 
holes of Ref.~\cite{Vines:2017hyw} reveals that it exhibits this symmetry and is 
broken by inclusion of general spin-induced multipole moments. 
We therefore conjecturally {\em define} the scattering amplitude of two Kerr black holes 
as the amplitude which realize the symmetry~\eqref{symmetry}. This definition fixes the 
Wilson coefficients; at 1PM order it recovers those of Ref.~\cite{Vines:2017hyw}.
It will be interesting to see if this definition holds at higher orders in~$G$.

This definition is consistent with the vanishing of $\delta^{(2,3)}$ for $C_2=1$.
Requiring that $\delta^{(3,3)}$ vanishes further sets $H_2=1$, which is also the 
value that leads to a Compton amplitude with good high-energy properties~\cite{Cucchieri:1994tx, Giannakis:1998wi, Cortese:2013lda}. 
Ref.~\cite{Vines:2016unv}  fixes $H_2=1$  for general bodies, not just for the Kerr black hole, by requiring that 
the form of the equation of motion is invariant under the change of SSC. 
See Ref.~\cite{Porto:2006bt, *Porto:2007tt, *Porto:2008tb,*Porto:2008jj} for an alternative perspective.
In contrast, string theory predicts state-dependent values for $H_2$, perhaps due to the spectrum containing 
more than a single massive higher-spin state~\cite{Giannakis:1998wi}.

Realization of the shift symmetry~\eqref{symmetry} at this order requires that 
$\delta^{(4,4)}, \delta^{(4,5)}$ and $\delta^{(4,6)}$ vanish.
Demanding this fixes $H_4=0$, which recovers the amplitude obtained in \cite{Chen:2021qkk} 
and, notably, leaves $H_3$ undetermined.  
The proposed symmetry~\eqref{symmetry} and the resulting values for Wilson coefficients are consistent 
with the $S_1^m S_2^{4-m}$ amplitude included in the ancillary file~\cite{anc}.

Requiring the presence of the symmetry \eqref{symmetry}, or that $\delta^{(5,4)}, \delta^{(5,5)}$ and $\delta^{(5,6)}$ vanish, 
determines $H_3 = 3/2$ and shows that $H_5$ is degenerate with $R^2$ terms at this order. It however leaves undetermined certain $R^2$ Wilson coefficients.
The combination appearing in the amplitude can be fixed by requiring that, as at lower powers of the spin, the amplitude does not grow faster than 
the spin-independent part at large $\sigma$, where we take ${\cal E}_{1, 2} \sim \sigma$ because of its momentum dependence. Equivalently, one may require that 
the classical part of the one-loop amplitude does not grow faster at high energies than the tree-level amplitude. The polynomials $\delta^{(5,i)}$ are 
then uniquely fixed, given below together with $\gamma^{(5,i)}$,
\begin{align}
&\gamma^{(5,1)}_{\rm Kerr}=
%%%%% begin : gammaKerrPaper[5,1]
16(7-13\sigma^2)
%%%%% end : gammaKerrPaper[5,1]
\,, & &\frac{1}{75}\delta^{(5,1)}_{\rm Kerr}=
%%%%% begin : deltaKerrPaperOver75[5,1]
24(1-4\sigma^2)
%%%%% end : deltaKerrPaperOver75[5,1]
\,,\nonumber\\
&\gamma^{(5,2)}_{\rm Kerr}=
%%%%% begin : gammaKerrPaperOver75[5,2]
32(11\sigma^2-5)
%%%%% end : gammaKerrPaperOver75[5,2]
\,, & &\frac{1}{75}\delta^{(5,2)}_{\rm Kerr}=
%%%%% begin : deltaKerrPaperOver75[5,2]
48 (2 + \sigma^2)
%%%%% end : deltaKerrPaperOver75[5,2]
\,,\label{KerrSpin5} \\
 &\gamma^{(5,3)}_{\rm Kerr}=
%%%%% begin : gammaKerrPaperOver75[5,3]
16(3\sigma^2-1)
%%%%% end : gammaKerrPaperOver75[5,3]
\,, & &\frac{1}{75}\delta^{(5,3)}_{\rm Kerr}=
%%%%% begin : deltaKerrPaperOver75[5,3]
8 (12 - 16 \sigma^2 + 7 \sigma^4)
%%%%% end : deltaKerrPaperOver75[5,3]
\,.\nonumber
\end{align}
These results are collected in an attached Mathematica text file~\cite{anc}. It is interesting to understand if the symmetry \eqref{symmetry} and the high-energy scaling are sufficient to determine all 
Wilson coefficients for all powers of the spin at ${\cal O}(G^2)$ and perhaps beyond when supplemented with other information 
such as tree-level matching~\cite{Vines:2017hyw}.

%%%%%%%%%%%%%%%%%%%%%%%%%%%%%  TABLE %%%%%%%%%%%%%%%%%%
\def\hs{ \hskip .5 cm }
\def\tdot{\hskip -2 pt \cdot \hskip -1.1 pt }
\def\qdot{\hskip -1. pt \cdot \hskip -1.2 pt }
\def\disp{\displaystyle}
\begin{table}[tb]
        \setlength{\tabcolsep}{1pt}
        \renewcommand{\arraystretch}{1}
{\footnotesize
        \begin{tabular}{lll}
                \hline
$ \vphantom{\raisebox{5pt}{A}} 
%%%%% begin : OperatorTable[1,1]
 \tfrac{1}{r^2}  {\bm L} \cdot {\bm S}_1 
%%%%% end : OperatorTable[1,1]
 \hskip .4 cm $
&
$ 
%%%%% begin : OperatorTable[2,1]
 \tfrac{1}{r^4}   ({\bm r} \cdot {\bm S}_1)^2
%%%%% end : OperatorTable[2,1]
 \hskip .8 cm $
&
$ 
%%%%% begin : OperatorTable[2,2]
 \tfrac{1}{r^2}  {\bm S}_1^2
%%%%% end : OperatorTable[2,2]
$
\\[3pt]
$ 
%%%%% begin : OperatorTable[2,3]
 \tfrac{1}{r^2} ( {\bm p} \cdot {\bm S}_1)^2 
%%%%% end : OperatorTable[2,3]
$
&
$ 
%%%%% begin : OperatorTable[3,1]
 \tfrac{1}{r^4}  ({\bm p} \cdot {\bm S}_1)^2 {\bm L} \cdot {\bm S}_1 
%%%%% end : OperatorTable[3,1]
$
& 
$
%%%%% begin : OperatorTable[3,2]
\tfrac{1}{r^6}  ({\bm r} \cdot {\bm S}_1)^2  {\bm L} \cdot {\bm S}_1
%%%%% end : OperatorTable[3,2]
 $
\\[3pt]
$ 
%%%%% begin : OperatorTable[3,3]
\tfrac{1}{r^4}  {\bm S}_1^2  {\bm L} \cdot {\bm S}_1
%%%%% end : OperatorTable[3,3]
$
&
$ 
%%%%% begin : OperatorTable[4,1]
\tfrac{1}{r^4}   ({\bm p} \cdot {\bm S}_1)^4
%%%%% end : OperatorTable[4,1]
$
&
$ 
%%%%% begin : OperatorTable[4,2]
\tfrac{1}{r^6}   ({\bm r} \cdot {\bm S}_1)^2    ({\bm p} \cdot {\bm S}_1)^2
%%%%% end : OperatorTable[4,2]
 $
\\[3pt]
$ 
%%%%% begin : OperatorTable[4,3]
\tfrac{1}{r^8}  ({\bm r} \cdot {\bm S}_1)^4  
%%%%% end : OperatorTable[4,3]
 $
&
$ 
%%%%% begin : OperatorTable[4,4]
\tfrac{1}{r^4}   ({\bm p} \cdot {\bm S}_1)^2 {\bm S}_1^2
%%%%% end : OperatorTable[4,5]
 $
& 
$ 
%%%%% begin : OperatorTable[4,5]
\tfrac{1}{r^6}   ({\bm r} \cdot {\bm S}_1)^2 {\bm S}_1^2
%%%%% end : OperatorTable[4,6]
 $
\\[3pt]
$ 
%%%%% begin : OperatorTable[4,6]
\tfrac{1}{r^4} {\bm S}_1^4
%%%%% end : OperatorTable[4,6]
 $
&
$
%%%%% begin : OperatorTable[5,1]
\tfrac{1}{r^6}  ({\bm p} \cdot {\bm S}_1)^4 {\bm L} \cdot {\bm S}_1 
%%%%% end : OperatorTable[5,1]
 $
&
$
%%%%% begin : OperatorTable[5,2]
\tfrac{1}{r^6}  {\bm S}_1^4  {\bm L} \cdot {\bm S}_1 
%%%%% end : OperatorTable[5,2]
$
\\[3pt]
$
%%%%% begin : OperatorTable[5,3]
\tfrac{1}{r^{10}}  ({\bm r} \cdot {\bm S}_1)^4 {\bm L} \cdot {\bm S}_1 
%%%%% end : OperatorTable[5,3]
\; $
&
$
%%%%% begin : OperatorTable[5,4]
\tfrac{1}{r^6}  ({\bm p} \cdot {\bm S}_1)^2  {\bm S}_1^2  {\bm L} \cdot {\bm S}_1 
%%%%% end : OperatorTable[5,4]
\hskip .3 cm $
&
$
%%%%% begin : OperatorTable[5,5]
\tfrac{1}{r^8}  ({\bm r} \cdot {\bm S}_1)^2  {\bm S}_1^2  {\bm L} \cdot {\bm S}_1 
%%%%% end : OperatorTable[5,5]
 $
\\[3pt]
$
%%%%% begin : OperatorTable[5,6]
\tfrac{1}{r^8}  ({\bm r} \cdot {\bm S}_1)^2 ({\bm p} \cdot {\bm S}_1)^2  {\bm L} \cdot {\bm S}_1  
%%%%% end : OperatorTable[5,6]
 \hskip -.7 cm \null $ 
& $\null$
& $\null$
\\[3pt]
                \hline
        \end{tabular}
}
    \caption{The Hamiltonian spin structures for the first five orders in $\bm{S}_1$. }
        \label{table:HamiltonianOnlyS1}
\end{table}
%%%%%%%%%%%%%%%%%%%%%%%%%%%%%%%%%%%%  TABLE %%%%%%%%%%%%%%%%%%%%%%%%

Ref.~\cite{Bern:2020buy} related directly the amplitude coefficients and 
the coefficients of a set of spin structures in the position-space Hamiltonian. 
Table~\ref{table:HamiltonianOnlyS1} includes those that depend only
on ${\bm S}_1$.  
We express the position-space Hamiltonian in terms of 2, 9, 18, 43, 86
${\bm S}^n$ structures, keeping both ${\bm S}_1$ and ${\bm S}_2$ for
$n = 1$, 2, 3, 4, 5, respectively.  These structures and their coefficients 
are included in Mathematica-readable files attached to this manuscript~\cite{anc}.
Using this Hamiltonian, which depends only on canonical variables, any 
bound or unbound  physical observable can be determined straightforwardly by 
solving Hamilton's equations. 
Inspection of the expression of the Hamiltonian~\cite{anc} reveals a novel feature: the 
presence of $\pmb{p}^{-2}$-dependent terms; interestingly after applying Schouten identities these singularities cancel from physical observables for Wilson coefficients corresponding to Kerr black holes. They also always cancel at tree level.

We compared the scattering angle through $S^4$ in the aligned-spin limit with the results of Ref.~\cite{Bini:2017pee, Vines:2018gqi} for Kerr black 
holes and found complete agreement. The explicit expressions used in this comparison are given in an ancillary file~\cite{anc}~\footnote{We thank Michele Levi for encouraging us to make the explicit values of the angle available.}. We also found that our scattering angle for generic bodies is consistent with a generalization of Ref.~\cite{Siemonsen:2019dsu} that departs from Kerr geometry~\cite{*[] [{. We thank Justin Vines for sharing the results with us.}] JV}. Interestingly, the coefficients of $R^2$ operators in the framework of Ref.~\cite{Siemonsen:2019dsu} are related to quadratic combinations of our $C_n$ coefficients.

Finally, let us comment on a special class of linear-in-curvature operators which also include factors of 
$\nabla_a \phi_s M^{ab}$, which may be thought of as the off-shell covariantization of the $p_a S^{ab}$.
An example is $(D_2/m^2)R_{abcd}\nabla_i\phi_sM^{ai}M^{cd}\nabla^b\phi_s$ at $S^2$.
The three-point matrix elements of all operators containing $\nabla_a \phi_s M^{ab}$ vanish in the classical limit, 
being proportional to the SSC $p_a S^{ab}=0$. 
The higher-point matrix elements of these operators are more subtle, and indeed explicit calculations reveal 
that these operators contribute nontrivially to classical observables.
Therefore, covariant operators that give the same three-point matrix elements upon using SSC 
do not necessarily give equal four- or higher-point matrix elements, nor is it guaranteed that the difference can be 
absorbed into curvature-square operators.
Interestingly, all contributions linear in $D_n$ to the classical amplitude are also 
proportional to $(H_2-1)$, which vanishes if $H_2=1$; this is however no longer the 
case for the nonlinear dependence on $D_n$.

\sectionskip
\Section{Conclusions.} 
In this paper we constructed the $\mathcal O(G^2)$ 2PM two-body Hamiltonian for general compact 
objects, including Kerr black holes. We did so by extracting it from a variety of new 
amplitudes computed in the field-theory approach of Ref.~\cite{Bern:2020buy}.
Our explicit results are included in the ancillary material~\cite{anc}. 
Here we comment on two new and unexpected features that we identified and require 
further investigation.

We encountered a larger number of operators with independent Wilson coefficients 
at each order in spin; they include all those of the worldline approach, and others that either 
have fixed coefficients or do not have an obvious counterpart in the worldline approach~\cite{JV}. 
%%%%%%%%
Effectively, every gauge-invariant structure of the four-dimensional classical Compton 
corresponds to an independent Wilson coefficient.  For
linear-in-curvature operators we find that apart from $C_i$ which
agree with the worldline perspective, the Wilson coefficients $H_i$
also have independent contribution, so through ${\cal O}(S^4)$ there
are three additional coefficients. We demonstrated this point by
explicitly computing amplitudes up to ${\cal O}(S^5)$.
%%%%
It is important to understand the origin of this additional freedom in our formalism, for example, 
whether it is a consequence of the unconstrained nature of the higher-spin field we use, and
whether it corresponds to astrophysical phenomena beyond the current worldline 
description.  
We expect that a categorization of all independent higher-spin interactions in both the worldline 
and field-theory approaches together with a systematic comparison of results for observables will help 
resolve these issues.

Based on our explicit results and the observation of a spin shift symmetry, we also conjectured 
that certain spin-dependent structures characterize the Kerr-black-hole interactions 
to all orders in spin and perhaps even to all orders in Newton's constant.
We proposed this together with the requirement that the amplitude grows no worse than the spin-independent part of ${\cal M}$ at
high energies as a field-theory {\em definition} of the Kerr black hole limit.
It would be important to understand the physical interpretation of the shift symmetry and whether these constraints properly single out an effective field theory 
that describes the Kerr black hole of general relativity and study their consequences.

\sectionskip
\Section{Acknowledgments.}                                                                 
We especially thank Justin Vines for his valuable insights and cross
checks for the aligned-spin scattering angle. We also thank Clifford Cheung, Aidan
Herderschee, Enrico Herrmann, Callum Jones, Julio Parra-Martinez, Ira
Rothstein, Michael Ruf, Trevor Scheopner, Chia-Hsien Shen and Mikhail
Solon for helpful discussions. We are grateful to Rafael Aoude, Kays
Haddad and Andreas Helset for sharing a preliminary version of their
paper~\cite{Aoude:2022trd} which contains a similar conjecture for black
holes, specifically matching our \eqn{KerrSpin5} after a change of
basis.  Z.B.  and D.K. are supported by the U.S. Department of Energy
(DOE) under award number DE-SC0009937. R.R. and F.T.~are supported by
the U.S.  Department of Energy (DOE) under award number~DE-SC00019066.
This project has received funding from the European Union's Horizon
2020 research and innovation program under the Marie Sklodowska-Curie
grant agreement No.~847523 'INTERACTIONS'.  We are also grateful to
the Mani L. Bhaumik Institute for Theoretical Physics for support.

\Section{Note added.}  After this paper was completed the very interesting
paper Ref.~\cite{Bautista:2022wjf} appeared, deriving Compton amplitudes by
solving the Teukolsky equation for scattering plane waves off Kerr
black holes.  Aligned-spin scattering angles are subsequently obtained
in the eikonal limit.  Our aligned-spin angles agree in the
overlap, though some differences in the Compton amplitudes remain to
be resolved.

%%%%%%%%%%%%%%%%%%%%%%%%%%

\bibliography{ref}

\end{document}